\documentclass[aps,prl,twocolumn,superscriptaddress,showpacs]{revtex4}
\usepackage{CJK}
\usepackage[Symbol]{upgreek}
\usepackage{bm}
\usepackage{amsmath}
\usepackage{graphicx}
\usepackage{mathptmx} 
\usepackage{endnotes}
\usepackage{color}
\usepackage{float}
\usepackage[FIGTOPCAP, hang, tight ]{subfigure}
\begin{document}
\begin{CJK*}{GB}{song}


\title{Phase Diagrams of Bi$_{1-x}$Sb$_{x}$ Thin Films with Different Growth Orientations}

\author{Shuang Tang}
\email[]{tangs@mit.edu}
\affiliation{Department of Materials Science and Engineering, Massachusetts Institute of Technology, Cambridge, MA
02139-4037, USA} 
\author{M. S. Dresselhaus}
\email[]{millie@mgm.mit.edu}
\affiliation{Department of Electrical Engineering and Computer
Science, Massachusetts Institute of Technology, Cambridge, MA
02139-4037, USA} \affiliation{Department of Physics, Massachusetts
Institute of Technology, Cambridge, MA 02139-4037, USA}

\date{\today}

\begin{abstract}
 A closed-form model is developed to evaluate the band-edge shift caused by quantum confinement for a two-dimensional non-parabolic carrier-pocket. Based on this model, the symmetries  and the band-shifts of different carrier-pockets are evaluated for Bi$_{1-x}$Sb$_{x}$ thin films that are grown along different crystalline axes. The phase diagrams for the Bi$_{1-x}$Sb$_{x}$ thin film systems with different growth orientations are calculated and analyzed.
\end{abstract}
\pacs{73.22.-f,73.61.At,73.61.Cw,73.90.+f,81.07.-b}
\maketitle
\end{CJK*}
Bismuth-Antimony alloys (Bi${}_{1-x}$Sb${}_{x}$) constitute a materials system that has been attracting considerable attention in recent decades. This class of materials is considered to be one of the best materials candidates for thermoelectrics and refrigeration in the cryogenic temperature range. In 1993, Hicks et al. pointed out that a low-dimensional thermoelectric material system has a remarkably enhanced figure of merit (ZT) compared to the same material in its three-dimensional form \cite{Hicks1996PhysicalReviewB10493,Hicks1993PhysicalReviewB12727,Hicks1993PhysicalReviewB16631}. Since then researchers have been investigating low-dimensional materials systems in search of higher ZT. For one-dimensional Bi${}_{1-x}$Sb${}_{x}$ materials, Rabin et al. have reported a systematic theoretical picture of Bi${}_{1-x}$Sb${}_{x}$ nanowires, as well as some experimental results \cite{Rabin2001AppliedPhysicsLetters81}. 
\begin{figure}
 \includegraphics[width=0.5\textwidth]{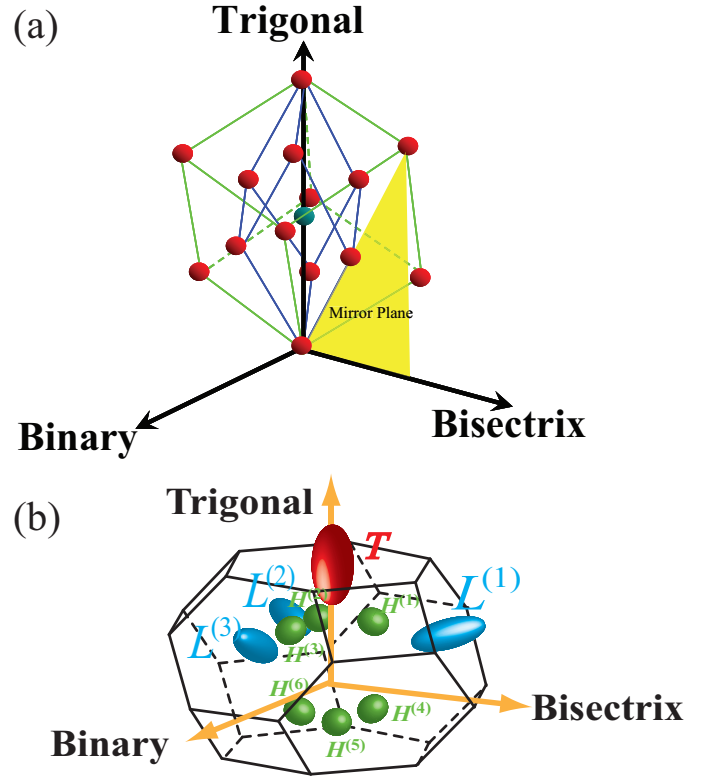}
\caption{Structure of  Bi${}_{1-x}$Sb${}_{x}$. (a) shows the rhombohedral lattice structure of bulk bismuth, bulk antimony and their alloys Bi$_{1-x}$Sb$_{x}$. There are 2 atoms in each unit cell. The $C_3$ symmetry trigonal axis, the $C_2$ symmetry binary axis and the $C_1$ symmetry bisectrix axis form a Cartesian coordinate system in three-dimensional space. The trigonal-bisectrix plane forms a mirror symmetry plane. (b) shows different carrier-pockets in the first Brillouin zone of bulk Bi$_{1-x}$Sb$_{x}$ are shown.} \label{structure}
\end{figure}

\begin{figure}
 \includegraphics[width=0.5\textwidth]{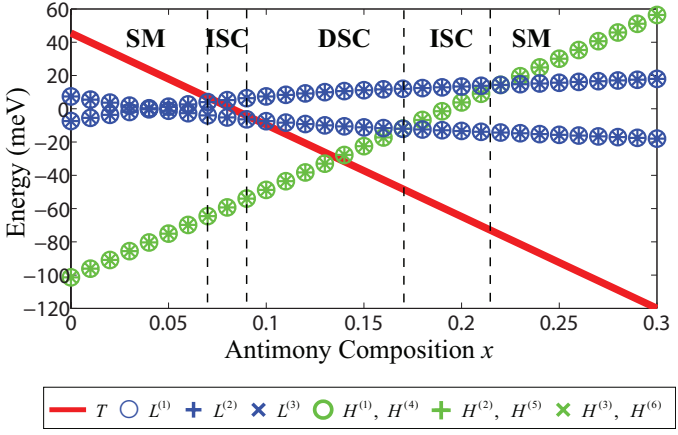}
\caption{Band-edge energies at different carrier-pockets vs. antimony composition $x$ for bulk  Bi${}_{1-x}$Sb${}_{x}$ \cite{Lenoir1996JournalofPhysicsandChemistryofSolids89}. The semi-metal (SM) phase regions, the indirect-semiconductor (ISC) phase regions and direct-semiconductor (DSC) phase region are labeled. The symbols for the various bands are indicated by the color and symbols in the bar below the plot.}\label{Exbulk}
\end{figure}

There are much richer phases in thin film Bi${}_{1-x}$Sb${}_{x}$ compared to bulk Bi${}_{1-x}$Sb${}_{x}$, because two more parameters are introduced: film thickness and growth orientation, which makes the research on  Bi${}_{1-x}$Sb${}_{x}$  thin films very interesting. Rogacheva et al. have synthesized Bi${}_{1-x}$Sb${}_{x}$ thin films that are grown normal to the trigonal crystalline axis \cite{Rogacheva2011}. Tang et al. have predicted the possibility for constructing a large variety of Dirac-cone materials based on the Bi${}_{1-x}$Sb${}_{x}$ thin film system \cite{Tang2011ArxivpreprintarXiv:1110.5391,Tang2011ArxivpreprintarXiv:1111.5525}. Thus, a phase diagram for Bi${}_{1-x}$Sb${}_{x}$ thin film system is strongly needed for experimentalists as a guide for sample synthesis for different research purposes. Such a phase diagram for Bi${}_{1-x}$Sb${}_{x}$ thin film system however has not yet been reported.  

In this present work, we develop electronic phase diagrams for Bi${}_{1-x}$Sb${}_{x}$  thin films that are synthesized normal to different crystalline axes for use in the cryogenic temperature range below the liquid nitrogen boiling point (77 K), and we study how the band-edges for different carrier-pockets change with antimony composition $x$, film thickness, and growth orientation. We firstly briefly review the relevant physical properties of bulk Bi$_{1-x}$Sb$_{x}$ alloys, and develop a closed-form model to evaluate the band-edge shift caused by the quantum confinement effect for a two-dimensional non-parabolic carrier-pocket. Based on this model we calculate the band-shifts and we  analyze the symmetries of different carrier-pockets for Bi$_{1-x}$Sb$_{x}$ thin films that are grown along the different principal crystalline axes. Finally, we calculate and analyze the phase diagrams for Bi$_{1-x}$Sb$_{x}$ thin film systems with different growth orientations.

We recall that bulk bismuth, bulk antimony and their Bi${}_{1-x}$Sb${}_{x}$ alloys  have a rhombohedral lattice structure with 2 atoms in each unit cell, where the $C_3$ symmetry trigonal axis, the $C_2$ symmetry binary axis, and the $C_1$ symmetry bisectrix axis that is perpendicular to the trigonal-binary plane, form a natural Cartesian coordinate system \cite{Schiferl1969JournalofAppliedCrystallography30}. The bisectrix axis and the trigonal axis form a mirror plane that bisects the whole rhombohedral lattice, as shown in Fig. \ref{structure}(a). In the first Brillouin zone of bulk Bi${}_{1-x}$Sb${}_{x}$, there are one $T$ point, three degenerate $L$ points (labeled as $L^{(i)}$, i=1,2,3) and six degenerate $H$ points (labeled as $H^{(j)}$, j=1,2...6) \cite{Jain1959PhysicalReview1518,Tichovolsky1969SolidStateCommunications927}, as shown in Fig.~\ref{structure}(b). The bottom of the conduction band is located at the three $L$ points. The top of the valence band can be located at the $T$ point, the three $L$ points or the six $H$ points, depending on antimony composition $x$. Figure~\ref{Exbulk} shows how the band-edges for different carrier-pockets change with antimony composition $x$ in bulk Bi${}_{1-x}$Sb${}_{x}$ alloys \cite{Lenoir1996JournalofPhysicsandChemistryofSolids89}, which leads to different phases: a semi-metal phase, an indirect-semiconductor phase and a direct-semiconductor phase. In the temperature range below 77 K, the band structure does not change notably with temperature.

For a Bi${}_{1-x}$Sb${}_{x}$ thin film, the three-fold symmetry of the $L$ points, the three-fold symmetry and the inversion symmetry of the $H$ points may be broken or reformed, and the normal-to-the-film quantum confinement effect of the carriers will increase the energy of the conduction band-edges and decrease the energy of the valence band-edges, and hence lead to different phase-transitions. The hole carrier-pockets at the $T$ point and the hole carrier-pockets at the six $H$ points are parabolically dispersive. The $T$-point and $H$-points quantum confinement effects relative to the bulk material can be modeled by the square-well model that has been used in Ref. \cite{Sandomirskii1967SovietJournalofExperimentalandTheoreticalPhysics101,Asahi1974PhysicalReviewB3347}. The valence band-edge at the $T$ point will be decreased by $h^2\cdot\alpha^{T}_{\perp[Film]}/{8\cdot l^2}$, and the valence band-edge at the $H^{(j)}$ point will be decreased by $h^2\cdot\alpha^{H^{(j)}}_{\perp[Film]}/{8\cdot l^2}$, where $l$ is the film thickness, $\alpha^{T}_{\perp[Film]}$ is the normal-to-the-film component of the $T$-point inverse-effective-mass tensor,  and $\alpha^{H^{(j)}}_{\perp[Film]}$ is the normal-to-the-film component of the $H^{(j)}$-point inverse-effective-mass tensor. As in previous reports, we can assume that the $T$-point and the $H^{(j)}$-point inverse-effective-mass tensors for Bi${}_{1-x}$Sb${}_{x}$ film and for bulk Bi${}_{1-x}$Sb${}_{x}$ are the same, i.e. $\alpha^{T}_{\perp[Film]}=\alpha^{T}_{\perp[Bulk]}$ and $\alpha^{H^{(j)}}_{\perp[Film]}=\alpha^{H^{(j)}}_{\perp[Bulk]}$.

However, this traditionally used model is not accurate to describe the quantum confinement effect for the  $L$-point  carriers. There is an electron carrier-pocket as well as a hole carrier-pocket at each of the $L^{(i)}$ points. The $L^{(i)}$-point conduction band-edge and valence band-edge energies are close, and therefore these bands are strongly coupled to each other, so that the dispersion relations are non-parabolic or perhaps even linear if Dirac cones are formed. The normal-to-the-film components of the inverse-effective-mass tensor $\alpha^{L}_{\perp[Film]}$ and the $L^{(i)}$-point band gap $E^L_{g[Film]}$ depend on the film thickness and are mutually coupled terms, which are both unknown. Thus, the simple model for calculating the $T$-point and $H^{(j)}$-point quantum confinement effects, which yield the band-edge shift from the normal-to-the-film component of the inverse-effective-mass tensor is not sufficient for evaluating the $L^{(i)}$-point band-edge shift or band gap increment. To calculate the $L^{(i)}$-point band-edge shift, we have developed a model starting from the three-dimensional two-band model that has been historically used in describing the non-parabolic $L$-point dispersion relations for bulk bismuth \cite{Lax1960PhysicalReviewLetters241}, which is:
\begin{equation} \label{GrindEQ__1_} 
\mathbf{p}\cdot \bm\upalpha^{L} \cdot \mathbf{p}=E(\mathbf{k})(1+\frac{E(\mathbf{k})}{E^{L}_{g} } ),                                                    
\end{equation} 
where $\bm\upalpha^{L}$ is the $L$-point inverse-effective-mass-tensor, and we assume here that $\bm\upalpha^{L}$ is the same for both the conduction band and the valence band within the context of a two-band model and strong interband coupling. Generally, the relation between $\bm\upalpha^L$ and $E_{g}^L$ around an $L$ point is described as
\begin{equation} \label{GrindEQ__2_} 
\bm\upalpha^L =\frac{2}{\hbar ^{2} } \frac{\partial ^{2} E(\mathbf{k})}{\partial (\mathbf{k} )^{2} } 
=\frac{1}{m_{0} } \cdot \mathbf{I}\pm \frac{1}{m_{0}^{2} } \frac{2}{E_{g}^L } \cdot \mathbf{p}^{2},                                    
\end{equation} 
under the $\mathbf{k}\cdot \mathbf{p}$ approximation \cite{Callaway1991,Rabin2001AppliedPhysicsLetters81,Rabin2004,Lin2001AppliedPhysicsLetters677,Lin2002AppliedPhysicsLetters2403}, where $\mathbf{I}$ is the identity matrix and $m_{0}$ is the free electron mass. We assume here that a two-band model also applies to Bi${}_{1-x}$Sb${}_{x}$ alloys, where the influence on $\bm\upalpha^L$ of adding antimony atoms up to a antimony concentration of \textit{x}=0.3 to bulk bismuth \cite{Callaway1991,Rabin2001AppliedPhysicsLetters81,Rabin2004,Lin2001AppliedPhysicsLetters677,Lin2002AppliedPhysicsLetters2403},  follows the relation:
\begin{equation} 
\label{GrindEQ__3_} 
\bm\upalpha^L (Bi_{1-x} Sb_{x} )=\frac{E^L_{g} (Bi)}{E^L_{g} (Bi_{1-x} Sb_{x})} \cdot (\bm\upalpha^L 
(Bi)-\frac{1}{m_{0}} \cdot\mathbf{I})+\frac{1}{m_{0}}\cdot 
\mathbf{I}.                    
\end{equation}
We further assume that Eq. (\ref{GrindEQ__3_}) is valid for Bi${}_{1-x}$Sb${}_{x}$ thin films as well, and hence we have 
\begin{widetext}
\begin{equation} 
\label{GrindEQ__3a_} 
\bm\upalpha^L _{[Film]}(Bi_{1-x} Sb_{x} )=\frac{E^L_{g[Bulk]} (Bi)}{E^L_{g[Film]} (Bi_{1-x} Sb_{x})} \cdot (\bm\upalpha^L_{[Bulk]} 
(Bi)-\frac{1}{m_{0}} \cdot\mathbf{I})+\frac{1}{m_{0}}\cdot 
\mathbf{I}.                    
\end{equation}
\end{widetext}
Both $\bm\upalpha^L _{[Film]}$ and $E^L_{g[Film]}$ of a Bi${}_{1-x}$Sb${}_{x}$ thin film differ from $\bm\upalpha^L _{[Bulk]}$ and $E^L_{g[Bulk]}$ in the bulk case, and they are both unknown at the beginning of the calculation. In Ref. \cite{Tang2011ArxivpreprintarXiv:1110.5391} and Ref. \cite{Tang2011ArxivpreprintarXiv:1111.5525} an iterative process is used to get the consistent $L$-point dispersion relations. In this present work, we are only interested in the band-edge energy shift, for which we have developed a closed-form model. We know that the coupling relation between $\bm\upalpha^L _{[Film]}$ and $E^L_{g[Film]}$ is described by Eq. (\ref{GrindEQ__3a_}), and because
\[
\frac{1}{m_0} \cdot (1-\frac{E^L_{g[Bulk](Bi)}}{E_{g[Film]}^L}) \cdot \mathbf{I}\ll \frac{E_{g[Bulk]}^L(Bi)}{E_{g[Film]}^L(Bi_{1-x}Sb_x)} \cdot \bm\upalpha_{[Bulk]}^L(Bi),
\]
Eq. (\ref{GrindEQ__3a_}) can be further simplified into
\begin{equation} \label{GrindEQ__5_} 
\bm\upalpha_{[Film]}^L (Bi_{1-x}Sb_{x})=\frac{\bm\upalpha_{[bulk]}^L(Bi)}{E_{g[Film]}^L (Bi_{1-x} 
Sb_{x} )} \cdot E_{g[Bulk]}^L(Bi). 
\end{equation}
The $L$-point band gap for the Bi${}_{1-x}$Sb${}_{x}$ thin film and the band-edge shift due to the quantum confinement effect are related by 
\begin{equation} 
\label{GrindEQ__4_} 
E_{g[Film]}^L(Bi_{1-x}Sb_x) =E_{g[Bulk]}^L(Bi_{1-x}Sb_x) +2\cdot \frac{h^{2} \alpha _{\perp[Film]}^L }{8\cdot l^{2}},
\end{equation}
where $E_{g[Bulk]}^L(Bi_{1-x}Sb_x)$ is the $L$-point band gap for bulk Bi${}_{1-x}$Sb${}_{x}$ with the same antimony composition $x$ as the film. From Eqs. (\ref{GrindEQ__5_}) and (\ref{GrindEQ__4_}), the $L$-point band gap for the Bi${}_{1-x}$Sb${}_{x}$ thin film can be solved to be
\begin{widetext}
\begin{equation}
\label{Eg}
E_{g[Film]}^L(Bi_{1-x}Sb_x)=\frac{E_{g[Bulk]}^L(Bi_{1-x}Sb_x)+\sqrt{(E_{g[Bulk]}^L(Bi_{1-x}Sb_x))^2+\frac{h^2}{l^2}\alpha^L_{\perp[Bulk]}(Bi)\cdot E_{g[Bulk]}^L(Bi)}}{2}
\end{equation}
\end{widetext}

\begin{figure*}
 \includegraphics[width=1\textwidth]{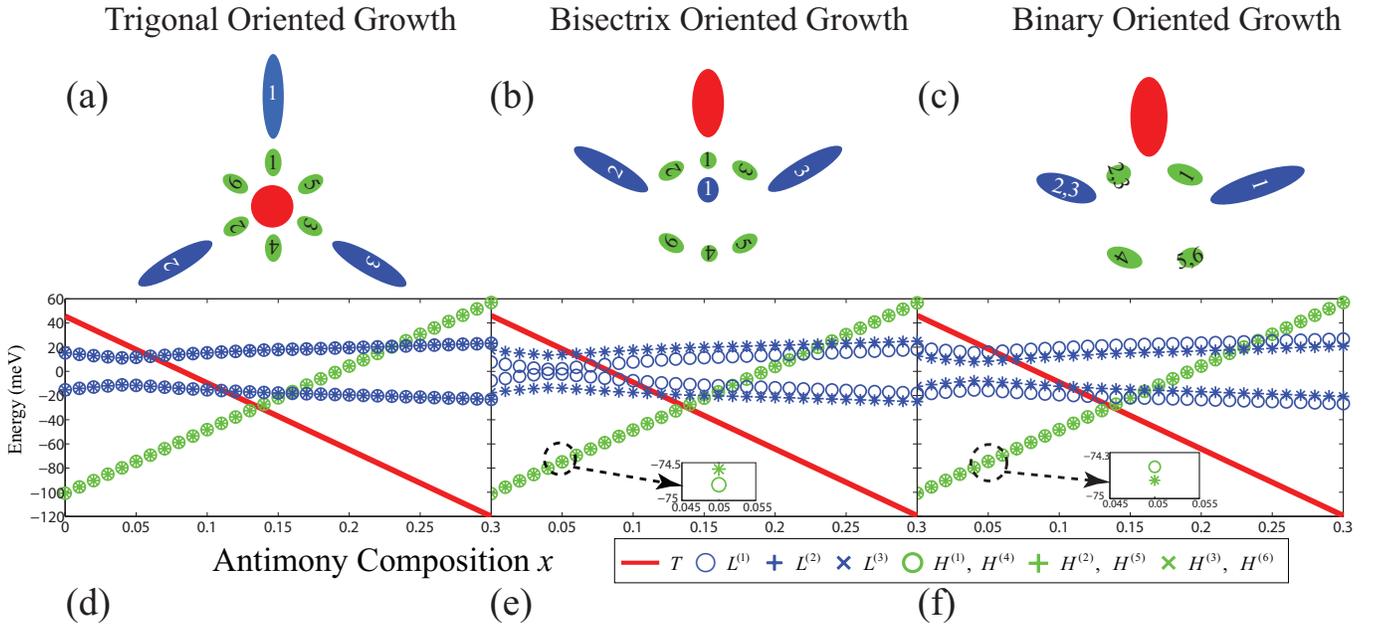}
\caption{ Schemes of how different carrier-pockets at the $T$ point (red), the three $L$ points (blue) and the six $H$ points (green) are projected on to the film plane for Bi${}_{1-x}$Sb${}_{x}$ thin films grown normal to (a) the trigonal axis, (b) the bisectrix axis and (c) the binary axis. How the band-edges for the different carrier-pockets change with antimony composition $x$ are calculated for Bi${}_{1-x}$Sb${}_{x}$ thin films grown normal to (d) the trigonal axis, (e) the bisectrix axis and (f) the binary axis, where 100 nm thick film systems are chosen as an example. The cases for other film thicknesses can be calculated in the similar way.}\label{Ex}
\end{figure*}

Figure \ref{Ex} (a)-(c) shows the schemes of how the various three-dimensional carrier-pockets are projected onto the film plane for different growth orientations denoted by the directions normal to the films. The calculated dependence of the band-edge energies for the different carrier-pockets in a Bi${}_{1-x}$Sb${}_{x}$ thin film against antimony composition are calculated in Fig. \ref{Ex} (d)-(f), where the 100 nm thick films of three different growth orientations are chosen as examples. Cases for other film thicknesses can be analysized in the similar way. 

Figures \ref{Ex}(a) and (d) show the case where the Bi${}_{1-x}$Sb${}_{x}$ thin film system is grown normal to the trigonal axis. From Fig. \ref{Ex}(a) we can see that the three $L$-point carrier-pocket projections have retained the three-fold symmetry. Moreover, the six $H$-point carrier-pocket projections have a six-fold symmetry and inversion symmetry. Due to the three-fold symmetry of the $L$-point projections, the conduction band-edges at the three points $L^{(1)}$, $L^{(2)}$ and $L^{(3)}$ are degenerate in energy, and so are the valence band-edges. 
Because of the six-fold symmetry and the inversion symmetry of the $H$-point projections, the valence band-edges at $H^{(1)}$, $H^{(2)}$, $H^{(3)}$, $H^{(4)}$, $H^{(5)}$, and $H^{(6)}$ are also all degenerate in energy. Compared to the bulk case shown in Fig.\ref{Exbulk}, the $T$-point valence band-edge and the $H$-points valence band-edges are not decreased significantly, because of the large effective masses normal to the film, of these carrier-pockets. However, the $L$-point band gaps are significantly increased, due to the large band-edge shift. The large band gaps and large band-shifts occur because of the small effective masses normal to the film of the $L$-point carrier-pockets. 

Figures \ref{Ex}(b) and (e) show the case where the Bi${}_{1-x}$Sb${}_{x}$ thin film system is grown normal to the bisectrix axis associated with the $L^{(1)}$ point. Because the bisectrix axis lies in the mirror plane in a bulk Bi${}_{1-x}$Sb${}_{x}$ lattice as shown in Fig. \ref{structure}, the $L^{(2)}$, the $H^{(2)}$ and the $H^{(5)}$ carrier-pocket projections are in mirror symmetry with respect to the $L^{(3)}$, the $H^{(3)}$ and the $H^{(6)}$ carrier-pocket projections, respectively, as shown in Fig. \ref{Ex}(b). The inversion symmetries of the $H$ points are still retained. These symmetries are reflected in Fig. \ref{Ex}(e), where the $L^{(2)}$-point and the $L^{(3)}$-point conduction (valence) band-edges are degenerate in energy, but are higher (lower) in energy than the $L^{(1)}$-point conduction (valence) band-edge. Meanwhile,The $H^{(j)}$ point is degenerate in energy with the $H^{(j+3)}$ point for j=1,2 and 3, due to inversion symmetry. The $H^{(1)}$-point and the $H^{(4)}$-point valence band-edges are lower in energy than the $H^{(2)}$-point, the $H^{(5)}$-point, the $H^{(3)}$-point and the $H^{(6)}$-point band-edges by a small amount due to the small effective masses  normal to the film . 

Figures \ref{Ex}(c) and (f) show the case where the Bi${}_{1-x}$Sb${}_{x}$ thin film system is grown normal to the binary axis associated with the $L^{(1)}$ point.  From Fig. \ref{Ex}(c) we can see that the $L^{(2)}$-point carrier-pocket projection and the $L^{(3)}$-point carrier-pocket projection overlap with each other and are different from the $L^{(1)}$-point carrier-pocket projection. The inversion symmetry of the $H$ points is retained. Meanwhile, the $H^{(2)}$-point carrier-pocket projection and the $H^{(3)}$-point carrier-pocket projection overlap with each other and are different from the $H^{(1)}$-point carrier-pocket projection; the $H^{(4)}$-point carrier-pocket projection and the $H^{(5)}$-point carrier-pocket projection overlap with each other and are different from the $H^{(6)}$-point carrier-pocket projection. The $H^{(j)}$ point is degenerate in energy with the $H^{(j+3)}$ point for j=1,2 and 3 because of the inversion symmetry. Thus, in Fig. \ref{Ex}(f), we can see that the conduction (valence) band-edges at the $L^{(2)}$ point and the $L^{(3)}$ point are degenerate with each other and  are lower (higher) in energy than the conduction (valence) band-edge at the $L^{(1)}$ point. The $H^{(1)}$-point and the $H^{(4)}$-point valence band-edges are higher in energy than the $H^{(2)}$-point, the $H^{(5)}$-point, the $H^{(3)}$-point, and the $H^{(6)}$-point band-edges by a small amount.

The phase of a Bi${}_{1-x}$Sb${}_{x}$ thin film can be decided by the relative positions in energy of the band-edges of the different carrier-pockets. When the $T$-point valence band-edge or any one of the six $H$-point valence edges is above all the three $L$-point conduction band-edges, this Bi${}_{1-x}$Sb${}_{x}$ thin film is in a semimetal phase. When the $T$-point valence band-edge and all the six $H$-point valence edges are below all the three $L$-point valence band-edges, this Bi${}_{1-x}$Sb${}_{x}$ thin film is in a direct-semiconductor phase. Otherwise, the Bi${}_{1-x}$Sb${}_{x}$ thin film is in an indirect-semiconductor phase
\begin{figure*}
 \includegraphics[width=1\textwidth]{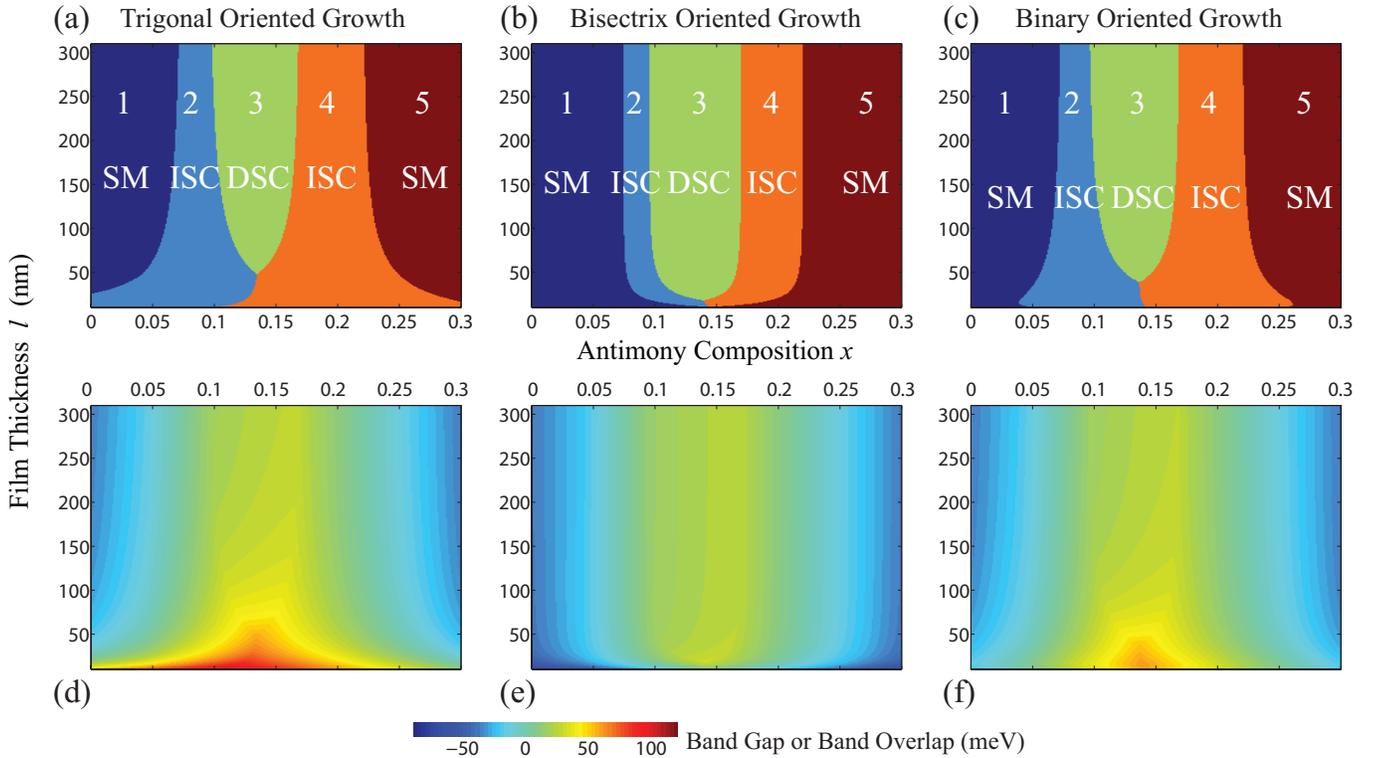}
\caption{Phase diagrams as a function of antimony composition $x$ and film thickness $l$ for the Bi${}_{1-x}$Sb${}_{x}$ thin film systems grown normal to (a) the trigonal axis, (b) the bisectrix axis and (c) the binary axis. The semi-metal (SM) phase regions, the indirect-semiconductor (ISC) phase regions and direct-semiconductor (DSC) phase regions are labeled for each case. Region 1 and Region 2 in each phase diagram are the regions where the top of the valence band is located at an $L$ point; Region 4 and Region 5 in each phase diagram are the regions where the top of the valence band is located at an $H$ point. The band gap (band overlap) value as a function of film thickness and antimony composition are shown in (d), (e) and (f), corresponding to (a), (b) and (c), respectively, where a positive value stands for the band gap of a direct- or indirect-semiconductor, while a negative value stands for the band overlap of a semi-metal. A zero value stands for a gapless state, which corresponds to a semimetal-semiconductor or semiconductor-semimetal phase transition.}\label{phase}
\end{figure*}

The phase diagrams of the Bi${}_{1-x}$Sb${}_{x}$ thin film system for different growth orientations are calculated and shown in Fig.~\ref{phase} as a function of film thickness and antimony composition. Figure \ref{phase} (a) shows the phase diagram for the Bi${}_{1-x}$Sb${}_{x}$ thin film system grown normal to the trigonal axis, as a function of film thickness $l$ and antimony composition $x$. Region 1 and Region 2 on the relatively bismuth-rich side stand for the semi-metal (SM) phase and indirect-semiconductor phase (ISC), respectively, for which the top of valence band is located at the $T$ point. It is observed that the critical value of antimony composition for the semimetal-semiconductor phase transition between Region 1 and Region 2 is smaller in a Bi${}_{1-x}$Sb${}_{x}$ thin film than in a bulk Bi${}_{1-x}$Sb${}_{x}$, and decreases monotonically with decreasing film thickness $l$ due to the quantum confinement effect. Region 3 stands for the direct-semiconductor (DSC) phase where both the bottom of the conduction band  and the top of the valence band are located at the three $L$ points. Because the direct band-gap at the $L$ points is on the order of 0 to ~$10^1$ meV for most cases, this region could be interesting for photonic research in the infrared wavelength range.  Region 4 and Region 5 on the relatively antimony-rich side stand for the indirect-semiconductor phase and semi-metal phase, respectively, for which the top of the valence band is located at the six $H$ degenerate points. The critical value of antimony composition for the semiconductor-semimetal phase transition between Region 4 and Region 5 is larger in a Bi${}_{1-x}$Sb${}_{x}$ thin film than in a bulk Bi${}_{1-x}$Sb${}_{x}$, and increases monotonically with decreasing film thickness $l$ due to the quantum confinement effect. Bi${}_{1-x}$Sb${}_{x}$ thin films in the semi-metal phase regions (Region 1 and Region 5) usually have higher carrier concentrations than the films in the semiconductor regions \cite{Tang2011ArxivpreprintarXiv:1110.5391}, so these semi-metal phase regions could be interesting for transport studies and electronic devices design. Bi${}_{1-x}$Sb${}_{x}$ thin films in the semiconductor regions (Region 2, Region 3 and Region 4) have small band gaps which are on the order of 0 to ~$10^1$ meV for most cases, and may have a higher Seebeck coefficient than their semi-metal counterparts. Therefore these regions could be interesting for cryogenic thermoelectrics and low-temperature refrigeration.  As the film thickness decreases, the semi-metal phase regions and the direct-semiconductor phase region shrink, while the indirect-semiconductor regions expand. At the triple-phase point, the $T$-point valence band-edge, the three $L$-point valence band-edges and the six $H$-point valence band-edges are all degenerate in energy, which indicates a very high density of states that could potentially yield a significantly enhanced Seebeck coefficient \cite{Mahan1996ProceedingsoftheNationalAcademyofSciences7436}.  The band gap (band overlap) value as a function of film thickness and antimony composition are shown in Fig. \ref{phase} (d) corresponding to Fig. \ref{phase} (a), where a positive value stands for the band gap of a direct- or indirect-semiconductor, while a negative value stands for the band overlap of a semi-metal.

Figure \ref{phase}(b) shows the phase diagram for the Bi${}_{1-x}$Sb${}_{x}$ thin film system grown normal to the bisectrix axis. In contrast to Fig. \ref{phase}(a), in Fig. \ref{phase}(b) the critical value of antimony composition for the semimetal-semiconductor phase transition between Region 1 and Region 2 is larger than that in a bulk Bi${}_{1-x}$Sb${}_{x}$, and this value of $x$ increases monotonically with decreasing film thickness $l$, while the critical value of antimony composition for the semiconductor-semimetal phase transition between Region 4 and Region 5 is smaller than that in a bulk Bi${}_{1-x}$Sb${}_{x}$, and decreases monotonically with decreasing film thickness $l$.  By comparison to Fig. \ref{phase}(a), the semi-metal phase regions (Region 1 and Region 5) in Fig. \ref{phase}(b) have expanded remarkably, as well as the direct-semiconductor phase region (Region 3), while the indirect-semiconductor phase regions (Region 2 and Region 4) in Fig. \ref{phase}(b) have shrunk remarkably. The differences of Fig. \ref{phase}(b) as compared to Fig. \ref{phase}(a) arrives because the $L$-point normal-to-the-film effective mass component along the bisectrix axis is much larger than that along the trigonal axis, which leads to a much stronger $L$-point quantum confinement effect. The band gap (band overlap) value as a function of film thickness and antimony composition are shown in Fig. \ref{phase} (e) corresponding to Fig. \ref{phase} (b).

Figure \ref{phase}(c) shows the phase diagram for the Bi${}_{1-x}$Sb${}_{x}$ thin film system grown normal to the binary axis. Figure \ref{phase}(c) seems to be intermediate between Fig. \ref{phase}(a) and Fig. \ref{phase}(b), because the effective mass component normal to the film along the binary axis has a value between the corresponding mass components along the trigonal axis and along the bisectrix axis. The band gap (band overlap) value as a function of film thickness and antimony composition are shown in Fig. \ref{phase} (f) corresponding to Fig. \ref{phase} (c). 

In conclusion, we have develped a closed-form model to calculate the band-edge shifts due to the quantum confinement effect for a non-parabolic carrier-pocket. We have examined the symmetry of the three two-dimensional $L$-point carrier-pockets and the six two-dimensional $H$-point carrier-pockets for Bi$_{1-x}$Sb$_{x}$ thin films with different growth orientations, and we have calculated the band-edge shifts for each case.  Finally, we have calculated and analysized the phase diagrams for  the Bi$_{1-x}$Sb$_{x}$ thin film systems with different growth orientations.

\begin{acknowledgments}
The authors acknowledge the support from AFOSR MURI Grant number FA9550-10-1-0533, sub-award 60028687. The views expressed are not endorsed by the
sponsor.
\end{acknowledgments}

\end{document}